\documentclass[12pt]{iopart}

\usepackage{iopams}  
\usepackage{graphicx}
\begin{document}

\title{Identified Hadron Production from the RHIC Beam Energy Scan}

\author{Lokesh Kumar (for the STAR Collaboration)}

\address{Physics Department, Kent State University, Kent, OH-44242, USA.}
\ead{lokesh@rcf.rhic.bnl.gov}
\begin{abstract}
 A current focus at RHIC is the Beam Energy Scan to
  study the QCD phase diagram -- temperature ($T$) vs. baryon chemical
  potential ($\mu_{B}$). 
The STAR experiment has collected data
for Au+Au collisions at $\sqrt{s_{NN}}=$ 7.7 GeV, 11.5 GeV, and 39 GeV
in the year 2010. 
We present midrapidity results on rapidity density, average transverse mass, and particle
ratios for identified hadrons from the STAR experiment. 
Collision dynamics are studied in the framework of
chemical and kinetic freeze-out conditions.
\end{abstract}


\section{Introduction}
The RHIC beam energy scan (BES)~\cite{bes} started in the year 2010. Its main
objectives are to search for the possible QCD phase
boundary
and to search for the possible QCD critical point in the QCD phase
diagram.
At low temperatures, the relevant degrees of freedom are expected to
be hadronic but at high temperatures, the quarks and gluons are the 
relevant degrees of freedom of the system. In the QCD phase diagram, 
at $\mu_{B}$ $\sim$ 0,
the transition from hadronic gas to quark gluon plasma (QGP) is expected to
be a crossover~\cite{lattice}. At large $\mu_{B}$, it is expected
to be a first order phase transition. The point where the first order
phase transition line ends is called the QCD critical point.

From the spectra and ratios of the produced hadrons, $(T,\mu_{B})$ 
space points on the QCD phase diagram can be obtained.
Once the
$(T,\mu_{B})$ space points are obtained, one can study various 
signatures of the possible QCD phase boundary and QCD critical
point. 
The data presented here are from Au+Au collisions at $\sqrt{s_{NN}}=$ 7.7,
11.5, and 39 GeV, for the
midrapidity ($|y|<0.1$) region using both the STAR Time
Projection Chamber (TPC) and the Time Of Flight (TOF)
detectors~\cite{tpctof}. 
The errors shown
in figures are statistical and systematic errors added in quadrature.

\section{Results and Discussions}
\subsection{Energy Dependence of Yields and Average Transverse Mass}
\begin{figure}
\begin{center}
\includegraphics[scale=0.29]{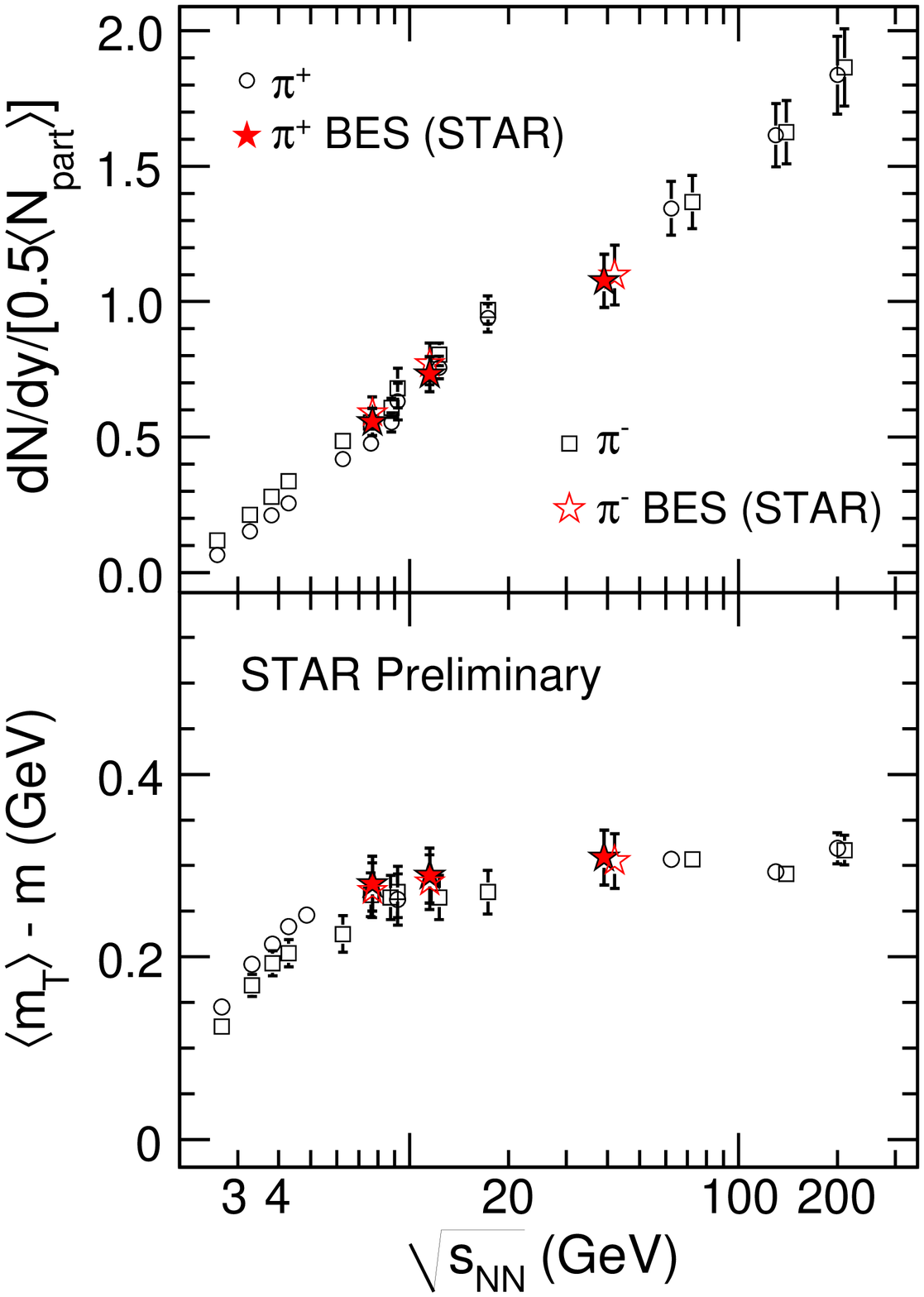}
\includegraphics[scale=0.29]{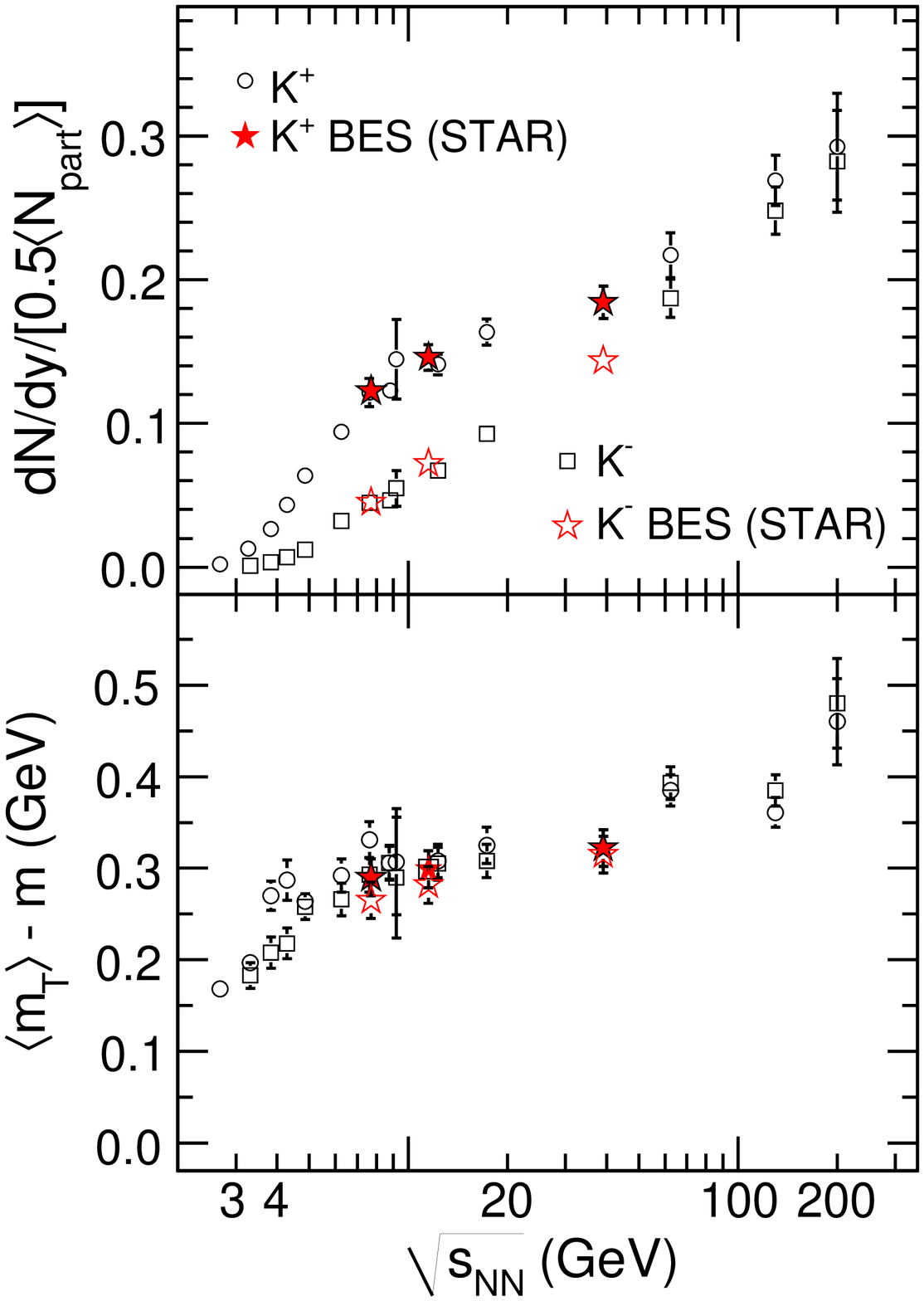}
\caption{\label{edep} (color online) Left: $dN/dy$ divided by $\langle N_{\rm{part}}
  \rangle$/2 
(top panel) and $\langle m_{T} \rangle - m$ (bottom panel) of $\pi^{\pm}$,
  plotted as a function of $\sqrt{s_{NN}}$ for central
  collisions at midrapidity. 
BES results (star symbols) are compared with published data 
 at RHIC~\cite{starpid}, SPS~\cite{sps}, and
 AGS~\cite{ags}. 
Right: Similar measurements for  $K^{\pm}$.  
}
\end{center}
\end{figure}

Figure~\ref{edep} shows the energy dependence 
of yields per
participating nucleon pair (top panels)
and the quantity $\langle m_{T}
\rangle - m $  (bottom panels) as a function of $\sqrt{s_{NN}}$ for
$\pi^{\pm}$ (left) and for $K^{\pm}$~(right). 
The pion yields for the BES data 
are corrected for weak decays feed-down and muon contamination using
STAR HIJING+GEANT, as done for previous STAR results~\cite{starpid}.
The BES results are consistent with the
published energy dependence trend. 
The yields per participating nucleon pair
increase with beam energy. 
The quantity $\langle m_{T} \rangle - m$,
where $m_{T} =\sqrt{p_{T}^{2} + m^{2}}$ and $m$ is the hadron mass, increases
with beam energy for lower energies, becomes almost constant for the region
covered by the BES data and then tends to increase towards the top RHIC
energies. This is an interesting observation in a scenario where the system is in a 
thermodynamic state. In that case, $\langle m_{T} \rangle - m$ can be
related to the temperature of the system and $dN/dy~(\propto log(\sqrt{s_{NN}}))$ may represent
entropy. Then this observation could reflect the
signature of a first order phase transition as proposed in 
ref.~\cite{vanhove}. 
However, other interpretations of the observed $\langle m_{T} \rangle - m$ are 
possible~\cite{bedanga}.

\subsection{Particle Ratios}
\begin{figure}
\begin{center}
\includegraphics[scale=0.31]{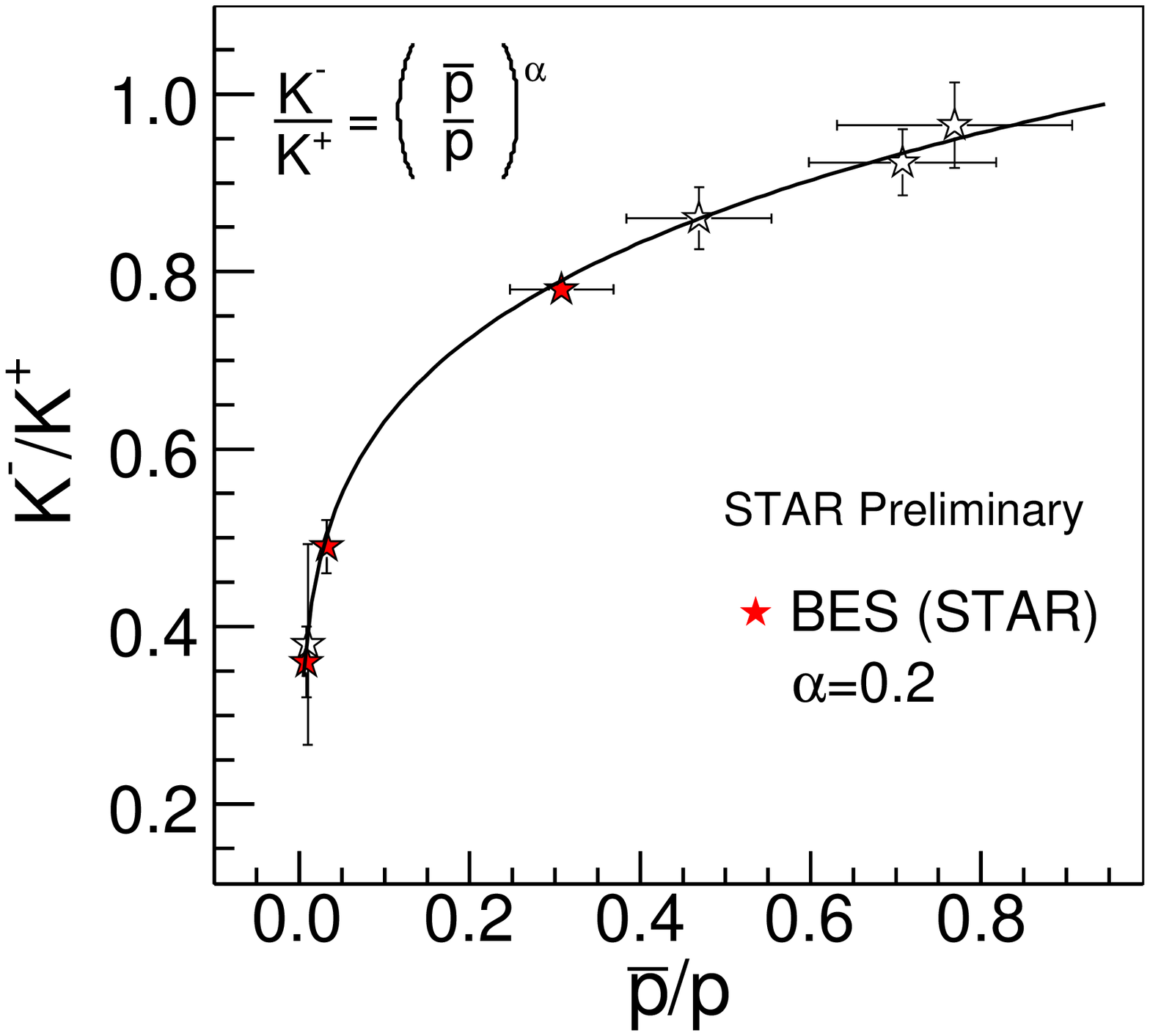}
\includegraphics[scale=0.30]{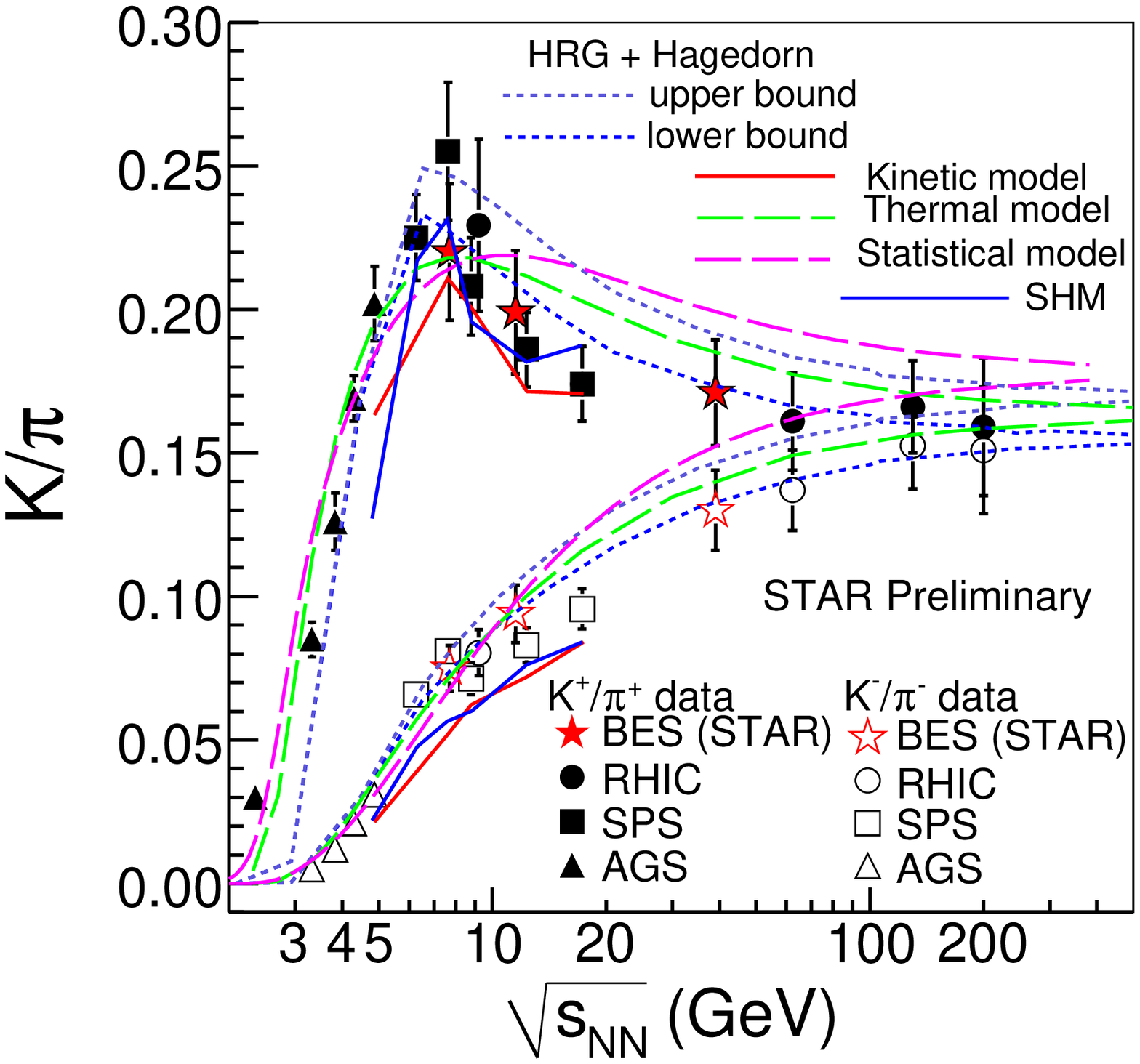}
\caption{\label{ratios} 
(color online) Left: Correlation of $K^{-}/K^{+}$ ratio with $\bar{p}/p$
ratio for central collisions at midrapidity for different energies. 
Right: Energy dependence of $K^{\pm}/\pi^{\pm}$
  ratio for central collisions at midrapidity.
 New results from BES
  data (star symbols) are compared to previously published
  results at RHIC~\cite{starpid}, SPS~\cite{sps}, and AGS~\cite{ags}, and are 
also compared with various
  theoretical model predictions~\cite{k2pimod}. 
}
\end{center}
\end{figure}

Figure~\ref{ratios} (left panel) shows the correlation of the
$K^{-}/K^{+}$ ratio with the $\bar{p}/p$ ratio for new measurements from
BES energies along with the published
results~\cite{starpid}. 
This could give information on how the kaon
production is related to the net-baryon density
($\bar{p}/p$ ratio).  The STAR $p$ and $\bar{p}$ yields are not feed-down corrected.
At lower energies, the kaon production is
dominated by the associated production which results in more $K^{+}$
production compared to $K^{-}$. Also the $\bar{p}/p$ ratio is much
less than
unity, indicating that there is large baryon stopping at the lower energies.
As we go towards higher energies, the pair production mechanism
starts to dominate and the ratios tend to become closer to
unity. 
The 
correlation between $K^{-}/K^{+}$ (representing net-strange chemical
potential, $\mu_{S}$) and $\bar{p}/p$ (representing net-baryon
chemical potential, $\mu_{B}$) ratios seems to
follow a power law behavior with $\alpha$=0.2 (represented by the
curve). In a hadron gas, the
relationship between $\mu_{S}$ and $\mu_{B}$ depends on the
temperature. In a particular case of $T$=190 MeV and $\mu_{B} <$ 500
MeV, these potentials follow the relation 
$\mu_{S} =(1/3) \mu_{B}$~\cite{kvsp}.

Recent theoretical calculations~\cite{cleyman2} suggest that the
maximum net-baryon density at freeze-out is attained at the lowest BES
energy of $\sqrt{s_{NN}} \sim $ 7.7 GeV. The maximum net-baryon density 
could also be related to the peak observed in the energy dependence of $K^{+}/\pi^{+}$
ratio at around $\sqrt{s_{NN}} \sim $ 7-8 GeV, as was observed 
by the NA49
experiment~\cite{sps}. 
This is sometimes referred to as the ``horn''. 
The $K/\pi$ ratio could also suggest 
the strangeness enhancement in heavy-ion collisions with respect to
the elementary collisions.
 Figure~\ref{ratios} (right panel) shows the
energy dependence of $K^{\pm}/\pi^{\pm}$ ratio for central collisions
at midrapidity. 
The BES results are in good agreement with the trend of
energy dependence established by the published measurements. 
The energy dependence of $K/\pi$ ratio seems
to be best explained using HRG+Hagedorn model~\cite{k2pimod}.

\subsection{Freeze-out Conditions}
\begin{figure}
\includegraphics[scale=0.26]{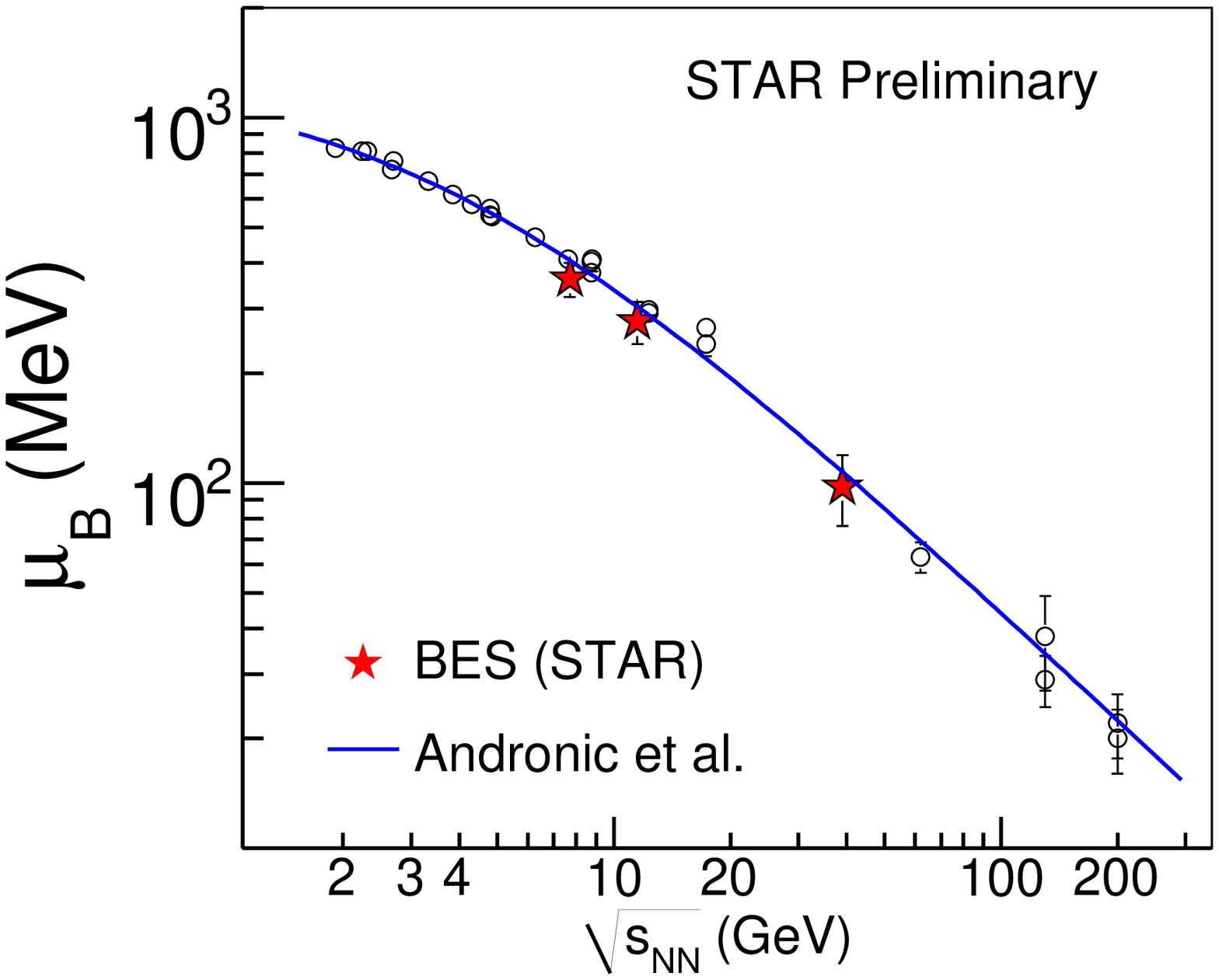}
\includegraphics[scale=0.27]{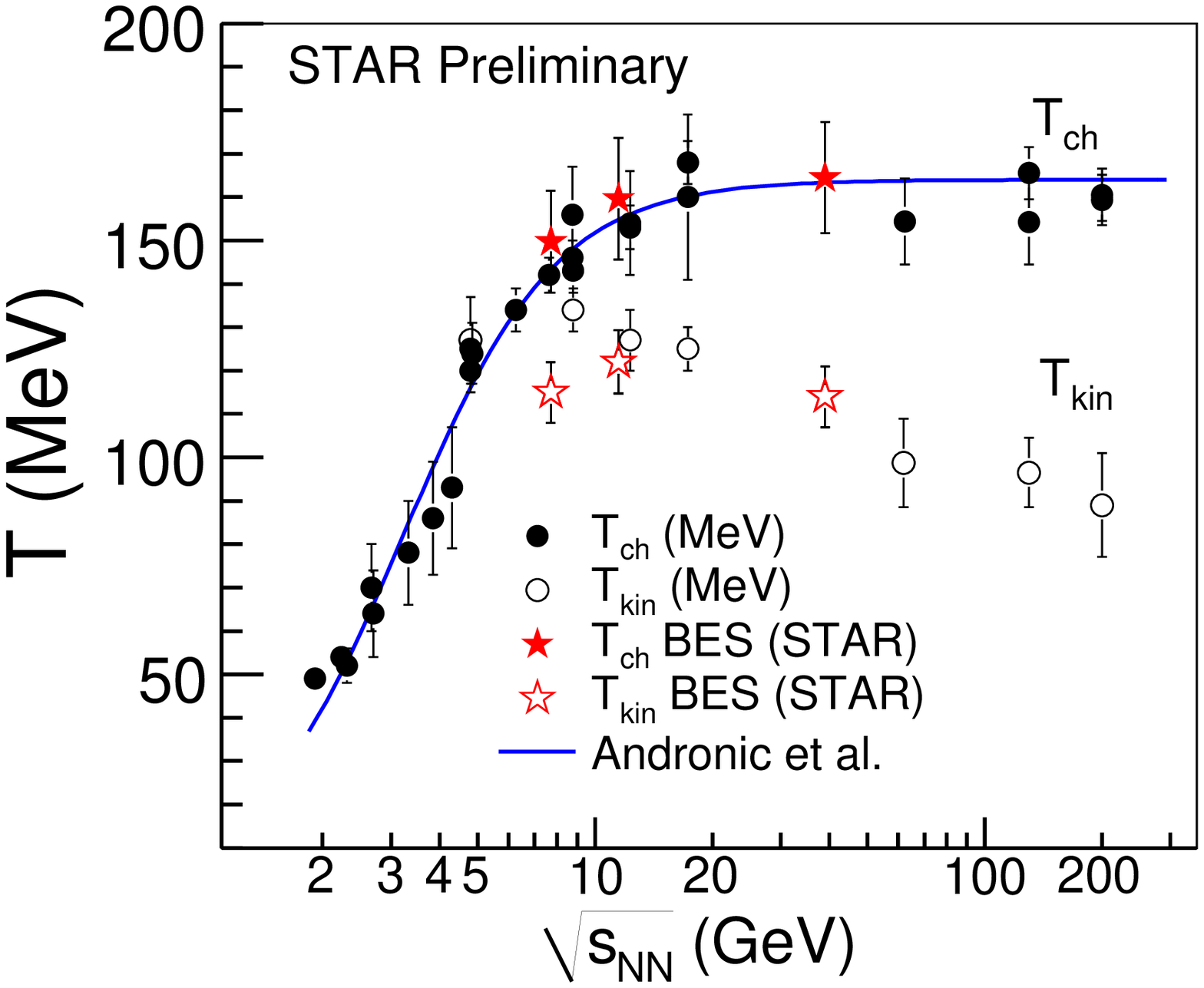}
\includegraphics[scale=0.26]{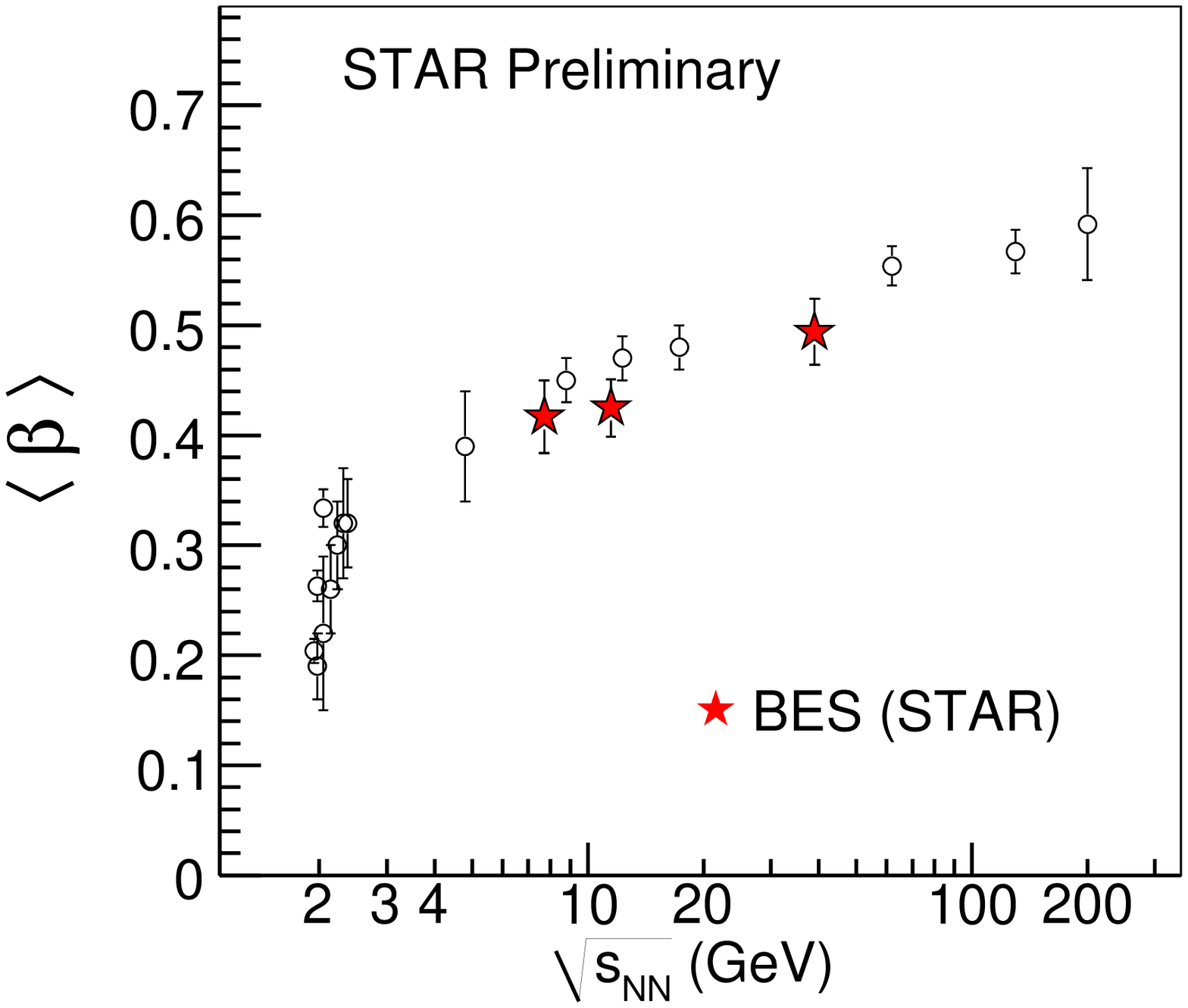}
\caption{\label{freezeout} 
(color online) Energy dependence of
 baryonic chemical potential (left), temperature ($T_{\rm{ch}}$,
 $T_{\rm{kin}}$) (middle),
 and average flow velocity (right). 
New results
 from BES data are shown with star symbols while other results are
 from ref.~\cite{starpid} and references therein.
 Lines represent the parametrization from ref.~\cite{anton}.
}
\end{figure}

The chemical and kinetic freeze-out conditions can be obtained by
comparing ratios and spectra of the produced particles with the
thermal equilibrium~\cite{starpid,chemfo} and  blast-wave
(BW)~\cite{starpid,bw} model calculations. The main parameters
extracted from these model comparisons are 
$T_{\rm{ch}}$ and 
$\mu_{B}$ for chemical freeze-out conditions, and 
$T_{\rm{kin}}$ and average
flow velocity $\langle \beta \rangle$ for kinetic freeze-out conditions.
Figure~\ref{freezeout}  
shows the energy dependence of freeze-out
parameters -
$\mu_{B}$ (left panel), $T_{\rm{ch}}$ and $T_{\rm{kin}}$ (middle
panel), and $\langle \beta \rangle$ (right panel), for central collisions.
The $\mu_{B}$ decreases with the beam energy. This is expected
since there are fewer net-baryons at midrapidity at higher energies
because of less stopping of baryons at midrapidity at higher
energies. The $T_{\rm{ch}}$ increases with energy
and saturates at the higher energies while $T_{\rm{kin}}$ 
decreases with beam energy after $\sqrt{s_{NN}} \sim$  7.7
GeV. 
The $\langle \beta \rangle$ increases
with beam energy.

In summary, bulk properties from the RHIC Beam Energy Scan are presented.
The results are in good agreement with the 
energy dependence trend established by the published measurements. 
The yields of identified hadrons increase with beam
energy.
The quantity $\langle m_{T} \rangle - m$ is almost constant for the BES
energies. 
The net-baryon density plays an important role at the lower
energies as demonstrated by the energy dependence of $K/\pi$ ratio and 
correlation between $K^{-}/K^{+}$ and $\bar{p}/p$ ratios. The new
measurements from the BES program extend the $\mu_{B}$ range covered by RHIC from 20--400 MeV.

\end{document}